\documentclass[12pt]{article}
\usepackage{amssymb}
\usepackage{graphics}
\usepackage{epsfig}

\DeclareFontFamily{U}{rsf}{}
\DeclareFontShape{U}{rsf}{m}{n}{
  <5> <6> rsfs5 <7> <8> <9> rsfs7 <10-> rsfs10}{}
\DeclareMathAlphabet\Scr{U}{rsf}{m}{n}

\catcode`\@=11
\def\@citex[#1]#2{%
\if@filesw \immediate \write \@auxout {\string \citation {#2}}\fi
\@tempcntb\m@ne \let\@h@ld\relax \def\@citea{}%
\@cite{%
  \@for \@citeb:=#2\do {%
    \@ifundefined {b@\@citeb}%
      {\@h@ld\@citea\@tempcntb\m@ne{\bf ?}%
      \@warning {Citation `\@citeb ' on page \thepage \space undefined}}%
      {\@tempcnta\@tempcntb \advance\@tempcnta\@ne%
      \@tempcntb\number\csname b@\@citeb \endcsname \relax%
      \ifnum\@tempcnta=\@tempcntb 
        \ifx\@h@ld\relax%
          \edef \@h@ld{\@citea\csname b@\@citeb\endcsname}%
        \else%
          \edef\@h@ld{\ifmmode{-}\else--\fi\csname b@\@citeb\endcsname}%
        \fi%
      \else
        \@h@ld\@citea\csname b@\@citeb \endcsname%
        \let\@h@ld\relax%
      \fi}%
    \def\@citea{,\penalty\@highpenalty\,}%
  }\@h@ld
}{#1}}

\def\@citeb#1#2{{[#1]\if@tempswa , #2\fi}}
\def\@citeu#1#2{{$^{#1}$\if@tempswa , #2\fi }}
\def\@citep#1#2{{#1\if@tempswa , #2\fi}}

\def\bcites{         
        \catcode`\@=11
        \let\@cite=\@citeb
        \catcode`\@=12
}

\def\upcites{         
        \catcode`\@=11
        \let\@cite=\@citeu
        \catcode`\@=12
}

\def\plaincites{      
        \catcode`\@=11
        \let\@cite=\@citep
        \catcode`\@=12
}

\newcount\hour
\newcount\minute
\newtoks\amorpm
\hour=\time\divide\hour by 60
\minute=\time{\multiply\hour by 60 \global\advance\minute by-\hour}
\edef\standardtime{{\ifnum\hour<12 \global\amorpm={am}%
        \else\global\amorpm={pm}\advance\hour by-12 \fi
        \ifnum\hour=0 \hour=12 \fi
        \number\hour:\ifnum\minute<10 0\fi\number\minute\the\amorpm}}
\edef\militarytime{\number\hour:\ifnum\minute<10 0\fi\number\minute}

\def\draftlabel#1{{\@bsphack\if@filesw {\let\thepage\relax
   \xdef\@gtempa{\write\@auxout{\string
      \newlabel{#1}{{\@currentlabel}{\thepage}}}}}\@gtempa
   \if@nobreak \ifvmode\nobreak\fi\fi\fi\@esphack}
        \gdef\@eqnlabel{#1}}
\def\@eqnlabel{}
\def\@vacuum{}
\def\marginnote#1{}
\def\draftmarginnote#1{\marginpar{\raggedright\scriptsize\tt#1}}
\def\draft{
        \pagestyle{plain}
        \overfullrule=2pt
        \oddsidemargin -.5truein
        \def\@oddhead{\sl \phantom{\today\quad\militarytime} \hfil
        \smash{\Large\sl DRAFT} \hfil \today\quad\militarytime}
        \let\@evenhead\@oddhead
        \let\label=\draftlabel
        \let\marginnote=\draftmarginnote
        \def\ps@empty{\let\@mkboth\@gobbletwo
        \def\@oddfoot{\hfil \smash{\Large\sl DRAFT} \hfil}
        \let\@evenfoot\@oddhead}
        \def\@eqnnum{(\theequation)\rlap{\kern\marginparsep\tt\@eqnlabel}%
        \global\let\@eqnlabel\@vacuum}  }

\def\section{\@startsection {section}{1}{\z@}{3.ex plus 1ex minus
 .2ex}{2.ex plus .2ex}{\large\bf}}
\def\subsection{\@startsection{subsection}{2}{\z@}{2.75ex plus 1ex minus
 .2ex}{1.5ex plus .2ex}{\bf}}

\def\appendix{{\newpage\section*{Appendix}}\let\appendix\section%
        {\setcounter{section}{0}
        \gdef\thesection{\Alph{section}}}\section}

\def\abstract{\if@twocolumn
\section*{Abstract}
\else 
\begin{center}
{\bf Abstract\vspace{-.5em}\vspace{0pt}}
\end{center}
\quotation
\fi}

\catcode`\@=12

\bcites

\catcode`\@=11
\def\theequation{\thesection.\arabic{equation}}
\@addtoreset{equation}{section}
\@addtoreset{footnote}{section}
\@addtoreset{footnote}{subsection}
\catcode`\@=12

\newcommand{\be}{\begin{eqnarray}}
\newcommand{\ee}{\end{eqnarray}}

\newcommand{\eqn}[1]{(\ref{#1})}
\def\Dslash{\,\,{\raise.15ex\hbox{/}\mkern-12mu D}}
\def\Dbarslash{\,\,{\raise.15ex\hbox{/}\mkern-12mu {\bar D}}}
\newcommand\delslash{\,\,{\raise.15ex\hbox{/}\mkern-9mu \partial}}
\def\delbarslash{\,\,{\raise.15ex\hbox{/}\mkern-9mu {\bar\partial}}}
\def\pslash{\,\,{\raise.15ex\hbox{/}\mkern-9mu p}}
\def\calDslash{\,\,{\raise.15ex\hbox{/}\mkern-12mu {\cal D}}}

\def\Tr{{\rm Tr}}

\newcommand{\Z}{{\bf Z}}

\newcommand{\R}{{\bf R}}
\newcommand{\RR}{{\mathbb R}}
\newcommand{\C}{{\bf C}}
\newcommand{\CC}{{\mathbb C}}
\newcommand{\PP}{{\mathbb P}}

\newcommand{\CP}{{\CC\PP}}

\newcommand\muo{\left.{\cal M}\right|_{U(1)}}
\newcommand{\mz}{\left.{\cal M}\right|_{z=0}}
\newcommand{\mw}{\left.{\cal M}\right|_{\omega=0}}

\begin{document}
\pagestyle{plain}
\setcounter{page}{1}
\newcounter{bean}
\baselineskip16pt

\begin{titlepage}

\begin{center}

\begin{flushright}
{\tt hep-th/0506022}\\
{\tt UT-Komaba/05-5}\\
{\tt MIT-CTP-3645} \\ 
{\tt  DAMTP-2005-52}
\end{flushright}

\vskip 2.5 cm {\Large \bf Reconnection of Non-Abelian Cosmic
Strings} \vskip 1 cm
{Koji Hashimoto${}^\dagger$ and David Tong${}^*$}\\
\vskip 1cm
{\sl ${}^\dagger$ Institute of Physics, University of Tokyo, 
Komaba\footnote{Permanent address}, Japan,}
 and \\
{\sl Center for Theoretical Physics, M.I.T., USA.}\\
{\tt koji@hep1.c.u-tokyo.ac.jp}\\
\vskip .3cm
{\sl ${}^*$ DAMTP, University of Cambridge, UK. \\
{\tt d.tong@damtp.cam.ac.uk}}

\end{center}

\vskip 0.5 cm
\begin{abstract}
Cosmic strings in non-abelian gauge theories naturally gain a
spectrum of massless, or light, excitations arising from their
embedding in color and flavor space. This opens up the possibility
that colliding strings miss each other in the internal space,
reducing the probability of reconnection. We study the topology of
the non-abelian vortex moduli space to determine the outcome of string
collision. Surprisingly we find that the probability of classical
reconnection in this system remains unity, with strings passing
through each other only for finely tuned initial conditions. We
proceed to show how this conclusion can be changed by symmetry
breaking effects, or by quantum effects associated to fermionic
zero modes, and present examples where the probability of
reconnection in a $U(N)$ gauge theory ranges from $1/N$ for
low-energy collisions to one at higher energies.
\end{abstract}

\end{titlepage}

\tableofcontents

\section{Introduction}

The recent revival of interest in cosmic strings is due to
developments on both observational and theoretical fronts. On the
observational side there is optimism that the next generation of
gravitational wave detectors, Advanced LIGO and LISA, will be able
to detect the characteristic signature of cosmic string loops as
they twist and whip into cusps \cite{dv,dv2}. Moreover, there exist
tantalizing gravitational lensing events, most notably CSL1 and
its companion images \cite{sazhin}, which point at the existence
of a cosmic string with tension $G\mu\sim 4\times 10^{-7}$.
Further spectroscopic analysis should reveal the nature of this
system in the near future.  Reviews of these developments can be
found in \cite{kibble,pollec}.

On the theoretical side, the advent of the ``warped throat'' in
realistic type II string compactifications has permitted a
resurrection of the (b)old idea \cite{witten} that cosmic strings
may be superstrings stretched across the sky \cite{one,dvil,cmp}. In
this modern guise, the cosmic string network may consist of both
fundamental strings, D-strings and wrapped D-branes. Reviews of
these recent stringy developments can be found in \cite{pollec}
and \cite{annetom}, while earlier work on cosmic strings from the
1980's and early 90's is summarized in \cite{book,review}.

Having admitted the theoretical possibility that cosmic strings
may be fundamental strings, the important question becomes: how
can we tell? As described in \cite{pollec,cmp}, there are two major
distinctions between fundamental strings and gauge theoretic
solitons in perturbative field theories. The first is the
existence of multi-tension string networks consisting of both
fundamental strings, and D-strings, and their various bound
states. The study of the dynamical scaling properties of such
networks is underway \cite{tye,edpaul}. The second distinguishing
feature of fundamental strings, and the one we focus on in this
paper, is their interaction cross-section. In the abelian-Higgs
model it is known that cosmic strings reconnect with unit
probability $P=1$ over a wide range of impact parameters. In contrast,
fundamental strings interact with probability $P\sim g_s^2$, where
the functional dependence on the angle of incidence and relative
velocity of the strings was determined in \cite{orig,jjp}. Similarly, 
D-strings may also pass through each other, reconnecting with probability 
$P<1$ \cite{jjp,amikoji}. 
A reduced probability for reconnection affects the scaling solution
for the string network, resulting in a larger concentration of
strings in the sky \cite{jst,sak,dv2}. Given enough data, it is not
implausible that one could extract the probability $P$ from
observation. For further work on cosmic superstrings, see \cite{more}.

Of course, it may be possible to engineer a gauge theory whose
solitonic strings mimic the behavior of fundamental strings.
Indeed, since strongly coupled gauge theories are often dual to
string theories in warped throats, this must be true on some
level. However, restricting attention to weakly coupled gauge
theories, we could ask if there exist semi-classical cosmic
strings which, like their fundamental cousins, reconnect with
probability $P<1$. A mechanism for achieving this was mentioned by
Polchinski in \cite{pollec}: construct a vortex with extra
internal bosonic zero modes. Two vortices could then miss each
other in this internal space. It was further noted that symmetry
breaking effects would generically give mass to these internal
modes, ruining the mechanism except in rather contrived models.

In this paper we present a model which realizes Polchinski's field
theoretic counterexample although, ultimately, in a rather
different manner than anticipated. Our model embeds the
cosmic string in a non-abelian $U(N)$ gauge theory, so that the
string may move in the internal color and flavor space. The formal
properties of vortices of this type have been studied extensively
of late (see \cite{amime,konishi,amime2,gsy}) although, until now, not
in the context of cosmic strings. Despite the presence of this
internal space we find, rather surprisingly, that cosmic strings
continue to reconnect with essentially unit probability, passing
through each other only for finely tuned initial conditions. This
result occurs due to the non-trivial topology in the interior
region of the two-vortex moduli space. However, we show that this
conclusion is changed when the internal modes gain a mass. Lifting
the vortex zero modes naturally leaves behind $N$ different cosmic
strings, each of which reconnects only with strings of the same
type while passing through other types of strings. The effect of
lifting the internal moduli space is therefore to reduce the
probability of reconnection at low-energies to $1/N$. At high
energies these effects wash-out, the strings may evolve into each
other, and reconnection again occurs with probability one.

In the next section we review the cosmic strings of interest and
explain how they gain an internal space of massless modes. In
section 3 we study the moduli space of two vortex strings and
argue that vortex strings only fail to reconnect for a set of
initial conditions of measure zero. In section 4 we examine
various further effects in the model, including quantum
dynamics on the vortex worldvolume, symmetry breaking masses
and fermionic zero modes, and show how these effects reduce the
probability of reconnection at low energies. We conclude with the
traditional conclusions.

\section{Non-Abelian Cosmic Strings}

In this paper we study the dynamics of cosmic strings living in a
non-abelian $U(N_c)$ gauge theory, coupled to $N_f$ scalars $q_i$,
each transforming in the fundamental representation,
\be
L=\frac{1}{4e^2}\Tr\,F_{\mu\nu}F^{\mu\nu}+\sum_{i=1}^{N_f}{\cal
D}_\mu q_i^\dagger{\cal D}_\mu q_i -\frac{\lambda e^2}{2} \Tr
\biggl(
\sum_{i=1}^{N_f}q_i\otimes q_i^\dagger - v^2\biggr)^2 \label{lag}\ee
When $N_c=N_f=1$, this is simply the abelian-Higgs model while,
for $N_c>1$, it is the simplest non-abelian generalization. The
Lagrangian \eqn{lag} enjoys a $SU(N_f)$ flavor symmetry, rotating
the scalars. In Section 4 we shall discuss the more realistic
situation in which this symmetry is softly broken, but for now let
us assume it remains intact. Moreover, we shall restrict attention
to the case $N_f=N_c\equiv N$\footnote{When $N_f<N_c$, the central
$U(1)$ remains unbroken and the theory does not admit vortices.
For $N_f>N_c$, the resulting cosmic strings are non-abelian
generalizations of semi-local vortices \cite{semi}.}. The theory
\eqn{lag} has a unique vacuum in which the scalars condense in the
pattern
\be q^a_{\ i}=v\delta^{a}_{\ i} \label{qia}\ee
Here $i=1,\ldots,N$ is the flavor index, while $a=1,\ldots N$ is
the color index. In what follows we will take our theory to be
weakly coupled by requiring the symmetry breaking scale $ev \gg
\Lambda$, the scale at which the non-abelian sector confines. The
vacuum \eqn{qia} has a mass gap in which the gauge field has mass
$m_\gamma^2\sim e^2v^2$, while the the scalars have mass
$m^2_q\sim \lambda e^2v^2$. The symmetry breaking pattern
resulting from this condensate puts the theory into what is
referred to as the ``color-flavor-locked'' phase, with
\be U(N_c)\times SU(N_f)\rightarrow SU(N)_{\rm
diag}\label{diag}\ee
The fact that the overall $U(1)\subset U(N_c)$ is broken in the
vacuum guarantees the existence of vortex strings characterized by
the winding $\int \Tr B = 2\pi k$ for some $k\in \Z$, where the
integral is over the plane transverse to the vortex. The tension
of this cosmic string is given by
\be T =2\pi v^2\, f(\lambda)\label{tension}\ee
where $f(\lambda)$ is a slowly varying function with  $f(1)=1$.
The width of the vortex core is given by $L\sim {\rm
max}(1/m_\gamma,1/m_q)$. The parameter $\lambda$ dictates the
behavior of multiple, parallel vortex strings: when $\lambda
> 1$, parallel vortex strings repel (as in a type II
superconductor) while, for $\lambda <1$, parallel strings attract
(type I superconductor). In both cases, the force is short ranged,
dying off exponentially away from the vortex core. For the
critical coupling $\lambda=1$, vortex strings feel no force and
multi-soliton solutions exist with parallel vortex strings sitting
at arbitrary positions.

Vortex solutions to the theory \eqn{lag} can easily be constructed
from solutions to the corresponding abelian theory. Let
$A^\star_\mu$ and $q^\star$ denote gauge and  Higgs profiles of
the abelian vortex solution. Then we can construct a non-abelian
vortex by simply embedding in the upper-left-hand corner thus:
\be q^a_{\ i}=\left(\begin{array}{cccc} q^\star & && \\
& v && \\ && \ddots & \\ &&& v \end{array}\right)\ \ \ \,\ \ \ \ \
(A_\mu)^a_{\ b}=\left(\begin{array}{cccc} A_\mu^\star &&&
\\ & 0 && \\ && \ddots & \\ &&& 0 \end{array}\right)
\label{embed}\ee
This is not the most general embedding. We can act on this
configuration with the $SU(N)_{\rm diag}$ symmetry preserved by
the vacuum to generate new solutions. Dividing out by the
stabilizing group, the space of vortex solutions is
\be \frac{SU(N)_{\rm diag}}{SU(N-1)\times U(1)}\cong \CP^{N-1}
\label{cpn}\ee
The existence of these internal, Goldstone modes, on the string
worldsheet means that, at low-energies, the string feels as if it
is propagating in a higher dimensional space $\R^{3,1}\times
\CP^{N-1}$. The size (K\"ahler class) of the internal $\CP^{N-1}$
space is given by \cite{amime}
\be r=\frac{\tilde{f}(\lambda)}{e^2} \label{r}\ee
where $\tilde{f}(\lambda)$ is, once again, a slowly varying
function of $\lambda$ and it can be shown that
$\tilde{f}(1)=2\pi$.

A few comments on the literature: vortex zero modes arising
through a mechanism of this type were previously studied in
\cite{navortex} although these authors considered unbroken gauge
symmetries, a situation which leads to further subtleties. The
term ``non-abelian strings'' is also used to refer to
simply-connected gauge groups broken to a discrete subgroup, often
giving rise to several types of cosmic string; see for example
\cite{spergel}. Such strings have rather different
properties from those considered here, such as the existence of
string junctions, and their dynamics shares features with
$(p,q)$ string networks \cite{cmp,edpaul}. Finally, we make no attempt
to embed our model in a viable GUT, preferring to concentrate
instead on the robust features of our vortex strings. A detailed
description of $SO(10)$ GUT strings can be found, for example, in
\cite{so10}.

The cosmological consequences of these internal modes mimic the
behavior of string moving in higher dimensions. The internal
currents carried by the string are akin to motion in the higher
dimensions and, through equipartition of energy, have the effect
of slowing down the motion of the strings in the three dimensions
of real space \cite{paul}. As we will explain in Section 4, in our
case these internal modes actually gain a small mass from quantum
effects and such currents cease to play a role over large times.

More important for the present discussion is the fact that the
internal space \eqn{cpn}, like the higher dimensions of string
theory, allows vortices to pass without interacting. To see this,
consider two abelian vortices in orthogonal $U(1)$ subgroups, but
at different points $x_1$ and $x_2$ in space,
\be q^a_{\ i}=\left(\begin{array}{cccc} q^\star(x_1) & && \\
& q^\star(x_2) && \\ && \ddots & \\ &&& v \end{array}\right)\ \ \
\,\ \ \ \ \ (A_\mu)^a_{\ b}=\left(\begin{array}{cccc}
A_\mu^\star(x_1) &&&
\\ & A_\mu^\star(x_2) && \\ && \ddots & \\ &&& 0 \end{array}\right)
\label{opposite}\ee
In this case, the two strings evolve independently and simply pass
through each other: no reconnection occurs. Of course, if the two
vortices instead lie in the same $U(1)$ subgroup, as in
\eqn{embed}, then vortices strongly interact and, as we review in
the next section, reconnect. The question is: what happens in the
intermediate situations when the two vortices lie in overlapping
$U(1)$ subgroups?

For fundamental strings moving in $d$ compactified dimensions of
string theory, one expects the classical probability of
reconnection to be suppressed by the geometric factor $l_s^d/V$,
where $V$ is the volume of the extra dimensions and
$l_s=\sqrt{\alpha^\prime}$ is the width of the string
\cite{dvil}. Together with the inherent
probability $P\sim g_s^2$ of fundamental string reconnection
\cite{orig,jjp}\footnote{In realistic string compactifications, potentials 
on the internal space suppress the geometrical suppression while the 
$g_s$ suppression remains \cite{cmp}.}, this leads to a reduced string 
cross-section, the net result of which is to increase the number of cosmic strings 
seen in the sky \cite{jst,sak,dv2}.

Naively one may imagine that our field theoretic model exhibits
similar behavior, with a critical separation in the internal space
distinguishing reconnecting strings from those that pass through
each other. This would then allow one to define a classical
probability $P$ of reconnection in this system by coarse graining
over the internal space. In the next section we turn to a detailed
study of this issue. We shall find that the vortices {\it always}
reconnect {\it unless} they lie in orthogonal subgroups. In some
sense, the situation of orthogonal subgroups \eqn{opposite} is
already the critical separation and the classical probability of
reconnection is $P=1$.

\section{The Classical Reconnection of Cosmic Strings}

In general the non-linear evolution of solitons is a difficult
question that requires numerical investigation. However, for the
low-energy scattering of cosmic strings we may reliably employ
analytical methods in which we restrict attention to the light
degrees of freedom describing the positions and internal
orientations of the two strings. This method, known as the {\it
moduli space approximation} \cite{manton}, has been successfully
applied to the abelian Higgs model where it was used to show that
vortex strings indeed reconnect \cite{edneil,modrecon}. Later
numerical simulations revealed that this result is robust, holding
for very high energy collisions \cite{matz}. These results
underpin the statement that gauge theoretic cosmic strings
reconnect with probability one. Here we present the moduli space
analysis for the non-abelian strings; it is to be hoped that a
similar robustness holds for the present result.

\subsection{Reconnection of $U(1)$ Strings}

Let us start by recalling how we see reconnection from the moduli
space perspective in the case of the abelian Higgs model
\cite{edneil,modrecon,amikoji}. One can reduce the dynamics of cosmic
strings to that of particles by considering one of two spatial
slices shown in Figure 1. The vertical slice cuts the strings to
reveal a vortex-anti-vortex pair. After reconnection, this slice
no longer intersects the strings, implying the annihilation of this
pair. Alternatively, one can slice horizontally to reveal two
vortices. Here the smoking gun for reconnection is the right-angle
scattering of the vortices at (or near) the interaction point, as
shown in Figure 1 (right). Such $90^{\rm o}$ degree scattering is a
requirement since, as is clear from the figure, the two ends of
each string are travelling in opposite directions after the
collision. By varying the slicing along the string, one can
reconstruct the entire dynamics of the two strings in this manner
and show the inevitability of reconnection at low-energies.

\newcommand{\onefigurenocap}[1]{\begin{figure}[h]
         \begin{center}\leavevmode\epsfbox{#1.eps}\end{center}
         \end{figure}}
\newcommand{\onefigure}[2]{\begin{figure}[htbp]
         \begin{center}\leavevmode\epsfbox{#1.eps}\end{center}
         \caption{\small #2\label{#1}}
         \end{figure}}
\begin{figure}[tbp]
\begin{center}
\epsfxsize=6.0in\leavevmode\epsfbox{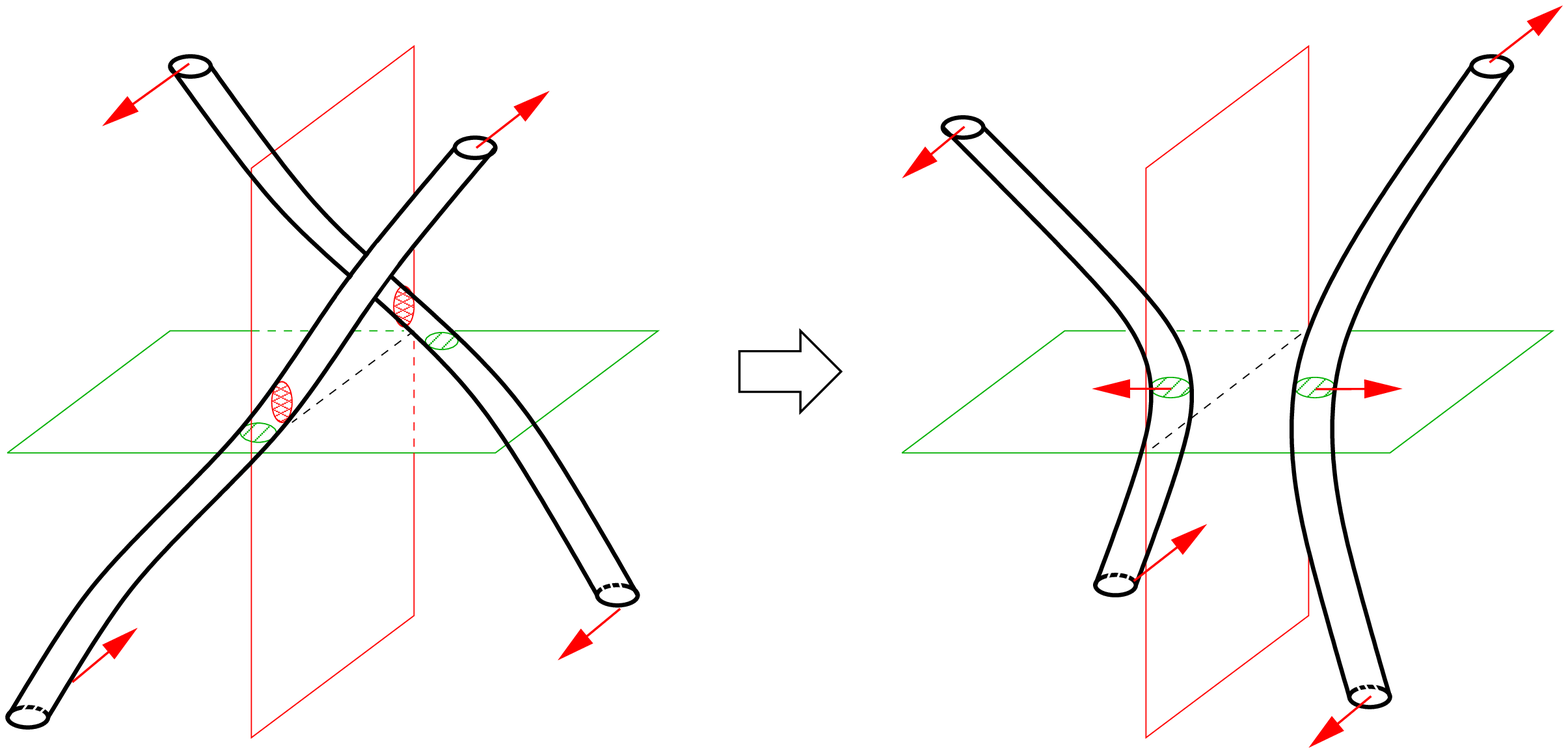}
\end{center}
\begin{center}
\begin{minipage}{13cm}
 \caption{\small 
The reconnection cosmic strings. Slicing vertically, one sees a
vortex-anti-vortex pair annihilate. Slicing horizontally, one sees
two vortices scattering at right angles.}
\end{minipage}
\end{center}
\label{fig:reconnect}
\end{figure}

Hence, reconnection of cosmic strings requires both the
annihilation of vortex-anti-vortex pairs and the right-angle
scattering of two vortices. While the former is expected (at least
for suitably slow collisions), to see the latter we must take a
closer look at the dynamics of vortices. At critical coupling
$\lambda=1$, the static forces between vortices cancel and we may
rigorously define the moduli space of solutions to the vortex
equations. The relative moduli space of two abelian vortices is simply
$\C/\Z_2$, where $\C$ is parameterized by $z$, the separation
between vortices, and $\Z_2:z\rightarrow -z$ reflects the fact
that the vortices are indistinguishable objects. This $\Z_2$ action means 
that the single valued coordinate on the moduli space is $z^2$, rather 
than $z$, an important point in what follows. While the metric
on this space is unknown\footnote{Various properties of the metric
on the vortex moduli space have been uncovered in
\cite{modvortex}.}, it is known to be smooth \cite{taubes}, looking 
like the snub-nose cone shown in Figure 2. The
motion of two particles at zero impact parameter goes up and over
the cone, as shown in the figure, returning down the other side.
This motion doesn't correspond to scattering by $180^{\rm o}$
(this would be coming back down the same side), but to $90^{\rm
o}$ scattering. This result does not depend on details of the
metric on the vortex moduli space, but follows simply from the
fact that, near the origin, the space is smooth and 
the single valued coordinate is $z^2$, rather than $z$.

\begin{figure}[tbp]
\begin{center}
\epsfxsize=13cm\leavevmode\epsfbox{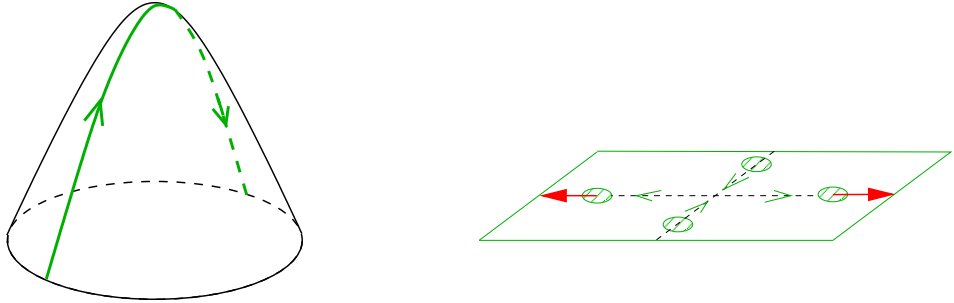}
\end{center}
\begin{center}
\begin{minipage}{13cm}
 \caption{\small
Right Angle Scattering: The view from the moduli space (on the
left) and in real space (on the right).}
\end{minipage}
\end{center}
\label{fig:ninety   }
\end{figure}
Before proceeding, we pass some well-known comments on the
validity of the moduli space approximation. The energies involved
in the collision should be small enough so as not to excite
radiation. In the present context, this means that the relative
velocity $\dot{z}$ of the vortices should satisfy $TL\dot{z}^2 \ll
m_\gamma, m_q$ where $L\sim \max (1/m_\gamma, 1/m_q)$ is the width
of the vortex string. For $\lambda \approx 1$ this translates into
the requirement that $\dot{z}^2\ll e^2$. Similarly, the angle of
incidence of the vortices, measured by $z^\prime$, the spatial
derivative of the separation along the string, should also satisfy
$z^{\prime 2} \ll e^2$. Finally, we should mention that the
description in terms of particle motion on the moduli space is a
little misleading, since a given slice of the string need not
follow a geodesic on the moduli space. (For example, waves
propagating along the string do not have this property). One
should talk instead in terms of the dynamics of the real line
$\R$, the spatial extent of the two strings, mapped into the
moduli space. The inevitability of reconnection then follows from
the single valued nature of $z^2$, together with the free motion
of the strings far from the interaction point \cite{amikoji}.

\subsection{The Moduli Space of Non-Abelian Vortices}

We would now like to repeat this analysis for the the non-abelian
vortices introduced in Section 2. For the vertical slice shown in
Figure 1, the abelian argument carries over. Our vortices have
only a single topological protector, $k = \int\Tr\, B/2\pi$, and a
vortex-anti-vortex pair may annihilate regardless of their mutual
orientation in the gauge group. One caveat is that a vortex and
anti-vortex in orthogonal $U(1)$'s (as in \eqn{opposite}) remains
a solution, albeit an unstable one. Therefore if we do not allow
fluctuations away from this ansatz, the vortex-anti-vortex pair
will pass right through each other without annihilating, in accord
with the statements in the previous section. This is our first
hint that reconnection will occur except for finely tuned initial
conditions.

To complete the argument of reconnection, we also need to study
when right-angle scattering occurs. For this we require a
description of the moduli space of multiple non-abelian vortices
in the critically coupled ($\lambda=1)$ theory \eqn{lag}. A
description of this space arising from modelling the system in
terms of string theoretic D-branes was presented in \cite{amime}.
The moduli space is presented in terms of an algebraic quotient
construction, related to the ADHM construction of the instanton
moduli space. We now review this construction.

The moduli space of $k$ vortices in $U(N)$ gauge theory is a
K\"ahler manifold with real dimension $2kN$ which we denote as
${\cal M}_{k,N}$. The construction of \cite{amime} presents this
space as a $U(k)$ symplectic quotient construction. We start with
a $k\times k$ complex matrix $Z$, and a $k\times N$ complex matrix
$\Psi$, subject to the constraint
\be [Z,Z^\dagger]+\Psi\Psi^\dagger=r \label{dterm}\ee
where the right-hand-side is proportional to the $k\times k$
identity matrix, and $r=2\pi/e^2$ as in \eqn{r}. The moduli space
${\cal M}_{k,N}$ is defined as the quotient of the solutions to
this constraint, where we divide by the $U(k)$ action
\be Z\rightarrow UZU^\dagger\ \ \ ,\ \ \ \Psi\rightarrow U\Psi
\label{bobby}\ee
The $U(k)$ action has no fixed points ensuring that, as in the
abelian case, the moduli space of vortices is smooth. Roughly
speaking, the eigenvalues of $Z$ correspond to the
positions\footnote{The eigenvalues of $Z$ are dimensionless and
correspond to the positions of the vortices multiplied by the mass
scale $v$.} of $k$ vortices, while the independent components of
$\Psi$ denote the orientations of these vortices in the internal
space. For example, when $k=1$, the scalar $Z$ decouples and
corresponds to the center of mass of the vortex, while $\Psi$
satisfies $|\Psi|^2=r$, modulo the $U(1)$ gauge action, which
reproduces the moduli space $\CP^{N-1}$ for a single vortex.

The manifold ${\cal M}_{k,N}$ has an $SU(N)_{\rm diag}\times
U(1)_R$ action. The former results from the symmetry \eqn{diag}
acting on the vortex; the latter is the rotational symmetry of the
plane. In the construction described above, the action is
\be SU(N)_{\rm diag}: \Psi\rightarrow \Psi V\ \ \ \ ,\ \ \ \
U(1)_R: Z\rightarrow e^{i\alpha}Z \ee
The algebraic quotient description of the vortex moduli space presented
here arises from studying vortices in a Hanany-Witten set-up
\cite{amime}. We stress that, despite the D-brane origin of this
construction, the resulting moduli space is that of field theoretic
vortices; indeed, in \cite{amikoji}, this framework was used to
elucidate the differences between abelian ($N=1$) vortex strings and
D-strings moving in vacua. Here we are interested in non-abelian 
($N\geq 2$) strings. To our knowledge, there is no field theoretic
derivation that the quotient space ${\cal M}_{k,N}$ coincides with the
vortex moduli space and such a proof would be desirable. In the
following we shall see that several key features of ${\cal M}_{k,N}$
correctly capture the behavior of vortex strings. Note however that the
space ${\cal M}_{k,N}$ naturally inherits a metric 
from the above construction; this does not coincide with the
metric on the moduli space of vortices (interpreted in terms of
solitons, it describes co-dimension two objects with long-range
polynomial tails). Thankfully, in what follows we will only
require topological information about ${\cal M}_{k,N}$.

\subsection{Reconnection of $U(2)$ Strings}

We first discuss the case of $k=2$ vortices in the $N_c=N_f=2$
gauge theory. Both $Z$ and $\Psi$ are $2\times 2$ matrices
(although for different reasons), and each suffers a $U(2)$ action
\eqn{bobby}. We project out the trivial center of mass motion of
the system by requiring $\Tr Z=0$ and, following \cite{kly}, use
the $U(2)$ action to place $Z$ in upper-triangular form. We write
\be
Z=\left(\begin{array}{cc} z & \omega \\ 0 & -z \end{array}\right)\ \ \ \ \ ,\ \ \ \ \ \
\Psi=\left(\begin{array}{cc} a_1 & a_2 \\ b_1 & b_2 \end{array}\right)
\ee
This does not fix all gauge degrees of freedom, but leaves a
surviving  $U(1)_1\times U(1)_2\times \Z_2 \subset U(2)$ gauge
symmetry acting as:
\be U(1)_1:\ U=\left(\begin{array}{cc} e^{i\phi} & 0 \\ 0 & 0
\end{array}\right)\ \ \ \ ,\ \ \ \ U(1)_2:\
U=\left(\begin{array}{cc} 0 & 0 \\ 0 & e^{i\phi}
\end{array}\right)\label{u1}\ee
under which $a_i$ transforms with charge $(1,0)$, $b_i$ with charge
$(0,1)$ and $\omega$ with charge $(1,-1)$. The coordinate $z$ is
neutral. Meanwhile, the $\Z_2$ action is
\be \Z_2:\ U=\frac{-1}{\sqrt{1+|\zeta|^2}}\left(\begin{array}{cc}
-1 & \bar{\zeta}
\\ {\zeta} & 1 \end{array}\right) \label{z2}\ee
with the parameter ${\zeta}=2z/\omega$. Finally, the constraints
arising  from \eqn{dterm} read
\be \sum_{i=1}^2|a_i|^2=r-|\omega|^2 \ \ \ ,\ \ \ \
\sum_{i=1}^2|b_i|^2=r+|\omega|^2\ \ \ ,\ \ \ \
 a_1\bar{b}_1+a_2\bar{b}_2=2\bar{z}\omega \label{3} \ee
Counting degrees of freedom, we have 6 complex parameters in $Z$
and $\Psi$ subject to two real constraints and one complex
constraint \eqn{3}, together with the two $U(1)$ actions \eqn{u1}.
This leaves us with a smooth moduli space $\tilde{\cal M}_{2,2}$
of 3 complex dimensions. (Recall that we have factored out the
center of mass degree of freedom so the full moduli space is
${\cal M}_{2,2}\cong \C\times \tilde{\cal M}_{2,2}$). The rest of
this section is devoted to studying this space.

\subsubsection*{\it The Asymptotic Regime}

To get a feel for the physical interpretation of the various
parameters, it is instructive to examine the regime of far
separated vortices with $z\gg r$. We have $|\omega|\sim 1/|z|$ and
the constraints \eqn{3}, combined with the $U(1)^2$ action
\eqn{u1}, restrict $a_i$ and $b_i$ to lie in independent
$\CP^1$'s, up to $1/|z|^2$ corrections. In this limit the $\Z_2$
action reads
\be \Z_2:
\left\{\begin{array}{ccc} z&\leftrightarrow & -z \\
a_i&\leftrightarrow& b_i
\end{array}\right.
\label{another}\ee
interchanging the position and orientation of the two vortices.
Thus, asymptotically, the moduli space is simply
\be \tilde{{\cal M}}_{2,2}\rightarrow\frac{\C \times \CP^1\times
\CP^1}{\Z_2} \ee

\subsubsection*{\it Two Submanifolds}

To continue our exploration of this space, it will prove useful to
seek out a couple of special submanifolds. These correspond to the
two extreme cases described in Section 2 in which we understand
that reconnection does/does not occur.

The first such submanifold corresponds to the situation
\eqn{embed} where the vortices lie in the same $U(1)$ subgroup. As
we mentioned in Section 2, such vortices must always scatter at
right-angles. We can impose this condition through the requirement
that the two orientation vectors lie parallel: $a_i\sim b_i$. We
will refer to this submanifold as $\muo\subset \tilde{\cal
M}_{2,2}$. By an $SU(2)_F$ action, we can choose a representative
point, say $a_2=b_2=0$. Then the constraints read
\be |a_1|^2=r-|\omega|^2 \ \ \ ,\ \ \ |b_1|^2=r+|\omega|^2\ \ \ ,\
\ \ \ a_1\bar{b}_1 = 2\bar{z}\omega \label{u1sub}\ee
This system was previously studied in \cite{kly}. On this
submanifold, $a_i$ and $b_i$ are both even under the $\Z_2$ action
\eqn{z2}, while $\omega$ and $z$ are odd: $(\omega, z)\rightarrow
-(\omega, z)$. The calculation of \cite{kly} shows that this
manifold is asymptotic to $\C/\Z_2$, with a smooth metric at the
origin, as depicted in Figure 2. Acting now with the $SU(2)_F$
action sweeps out a $\CP^1$ at each point, leaving us with a space
which is topologically\footnote{Topologically $\C/\Z_2\cong\C$. We
keep the former to emphasize that this description also captures
the asymptotic metric on the space.}
\be \muo\cong \C/\Z_2\times \CP^1 \ee
Note that as $z\rightarrow 0$, the $\CP^1$ does not vanish. In
this limit the equations \eqn{u1sub} are solved by $|\omega|^2=r$
and $a_i=0$, while $b_i$ parameterize a $\CP^1$ with K\"ahler
class $2r$.

 Let us now turn to the submanifold describing vortices in
orthogonal $U(1)$ subgroups as in \eqn{opposite}. The vortices
should now pass through each other without interacting. This
submanifold is defined by the requirement $\omega=0$, and\footnote{
This situation is similar to the theory describing two D-strings, 
in which $r=0$ and there is no $\Psi$ field, forcing $\omega=0$ \cite{amikoji}.}
we will
refer to it as $\mw$. The first two equations in \eqn{3} tell us
that $a_i$ and $b_i$ each define a point on $\CP^1$, while the
third equation, which reads $a_i\bar{b}_i=0$, requires these
points to be antipodal. Again, acting with the $SU(2)_F$ symmetry
then sweeps\footnote{Generically the orbits of the $SU(2)_F$
action on $\tilde{M}_{2,2}$ are three dimensional. They degenerate
to two dimensional orbits on $\muo$ and $\mw$.} out a $\CP^1$. We
still have to divide out by the $\Z_2$ gauge action which this time
acts as in \eqn{another}, exchanging $z\leftrightarrow -z$ and, at
the same time, mapping antipodal points on $\CP^1$. We therefore
have,
\be \left.{\cal M}\right|_{\omega=0} \cong \frac{{\C}\times
{\CP}^1}{\Z_2} \label{mw}\ee
But what happens at the origin? If we set $\omega=z=0$ then $\Psi$
feels the full force of the restored $U(2)$ gauge symmetry,
resulting in a unique solution to the constraints \eqn{dterm}.
This means that when the vortices live in orthogonal $U(1)$'s, as
in \eqn{opposite}, the internal space collapses as they approach
each other! $\mw$ can be thought of as the cone over $({\bf
S}^1\times \CP^1)/\Z_2$ (which can alternately be described as the
non-trivial ${\bf S}^2$ bundle over ${\bf S}^1$, or as the connect
sum $\RR\PP^3 \# \RR\PP^3$). Note that the submanifold $\mw$ is
singular at $z=0$. This is an artifact of restricting attention to
this subspace; the full manifold $\tilde{\cal M}_{2,2}$ should be
smooth at the point $\omega=z=0$.

How can we understand the result that the internal space collapses
at the origin of $\mw$ from the perspective of the soliton
solutions? In fact, it is rather simple to see. The solutions of
the form \eqn{opposite} generically transform non-trivially under
the $SU(2)_{\rm diag}$ vacuum symmetry, sweeping out the $\CP^1$
internal space. However, as $x_1\rightarrow x_2$ (corresponding to
$z\rightarrow 0$), the relevant part of $A_\mu$ and $q$ approaches
the unit matrix, and the $SU(N)_{\rm diag}$ symmetry no longer
acts. Two coincident vortices in orthogonal $U(1)$ sectors have no
internal space! This is one of the important points that allows
for reconnection to generically occur in this model.

\subsubsection*{\it Reconnection and the Origin of Moduli Space}

Having determined the topology of these two submanifolds, let us
now examine whether reconnection takes place on each. We start
with $\muo$. Here the argument proceeds as for the abelian case:
the submanifold $\C/\Z_2$ is a smooth cone, as shown in Figure 2,
with $z^2$ the single valued coordinate at the tip of the cone.
This ensures that any trajectory hitting the tip of the cone
results in right angle scattering. The $\Z_2$ does not act on the
internal space $\CP^1$ and it plays no role in the discussion of
reconnection.

What about the space $\mw$, describing orthogonal vortices? Here
the issue is somewhat clouded by the singularity at the center of
the space. However, consider first the resolved space where the
$\CP^1$ does not degenerate at the origin. Since the $\Z_2$ action
has no fixed points, such a manifold is smooth. In contrast to the 
previous case, a trajectory
through the origin at $z=0$ now corresponds to vortices passing straight 
through each other; there is no right angle scattering. The reason for this 
is that the $\Z_2$ gauge symmetry does not act only on $\C$ but
also on the internal space; it exchanges both the positions and the identities 
of the particles. This means that, near the origin, $z$ is the
single valued coordinate rather than $z^2$ and right-angle scattering does not
occur. The true motion in $\mw$, in which the $\CP^1$ degenerates, can be 
thought of as the limiting case of this discussion. The need to take this
degenerative limit is necessary since, as we shall presently see, 
$\omega=0$ corresponds to the only case of no reconnection; if the moduli 
space $\mw$ were not described by this singular limit then, 
by continuity, vortices in the neighborhood of
$\omega=0$ should also pass through each other.

Let us now show that, as promised, vortices of arbitrary orientation 
always scatter at right angles unless $\omega=0$. As we have seen above, 
the key to this lies in the $\Z_2$ action \eqn{z2}. In particular, we are 
interested in this action as the vortices approach each other
and $z\rightarrow 0$. Then we see that, {\it provided} $\zeta=2z/\omega
\rightarrow 0$, the $\Z_2$ action on $\Psi$ and $\omega$ can be
absorbed in the $U(1)^2$ gauge transformations, leaving only
$z\rightarrow -z$. In other words, for all $\omega\neq 0$, the
single valued coordinate near the origin is $z^2$ rather than $z$. 
Using the general arguments described above, this implies right-angle 
scattering and reconnection of cosmic strings.

To complete the argument, we need to make sure that restricting to
$z=0$ where the vortices coincide defines a complete submanifold.
Let us denote this as $\mz$. It may be defined in a coordinate
independent manner as the fixed locus of the $U(1)_R$ action
(since the resulting action on $\omega$ is gauge equivalent to a
$SU(2)_F$ rotation). When $z=0$, the $\Z_2$ action \eqn{z2} can be
absorbed into the axial combination of the $U(1)_1\times U(1)_2$
gauge symmetry \eqn{u1} and $\mz$ can be thought of as the
resolution of the $\Z_2$ fixed point. What is $\mz$? Upon setting
$z=0$, the constraints \eqn{3} read
\be |a_1|^2+|a_2|^2+|\omega|^2=r \ \ \ , \ \ \
|b_1|^2+|b_2|^2-|\omega|^2=r\ \ \ ,\ \ \  a_i\bar{b}_i= 0
\label{6} \ee
where we must still quotient by the $U(1)^2$ action \eqn{u1}.  The
first constraint, together with the $U(1)_1$ action, defines a
copy of $\CP^2$ (although it doesn't inherit the round
Fubini-Study metric). For a generic point $p\in\CP^2$, the second
and third constraints in \eqn{6} determine $b_i$ uniquely up to a
phase, which is gauged away by $U(1)_2$. If this were true
globally, we would have
\be \mz\cong {\CP}^2 \label{mz}\ee
However, there are two exceptional points. Firstly, when $a_i=0$
and $|\omega|^2=r$, the $b_i$'s parameterize a $\CP^1$ rather than
a point. We have $\muo\cap\mz\cong\CP^1$. Secondly, when
$\omega=0$, the full $U(2)$ gauge symmetry is restored and the
$a_i$'s parameterize a point rather than a $\CP^1$; we have
$\mw\cap\mz\cong\{0\}$. In fact, it is a rather cute fact that
after making these two adjustments, \eqn{mz} remains correct! To
see this, we may think of $\mz$ as a fiber over the interval
$|\omega|\in [0,\sqrt{r}]$. The phase of $\omega$ shrinks to zero
at each end due to the action of the gauge symmetry. In the middle
of the interval, $a_i$ and $b_i$, modulo the constraints \eqn{6}
and the $U(1)^2$ gauge action, define a ${\bf CP}^1$. The phase of
$\omega$ is fibered over this to yield a ${\bf S}^3$ (to see this,
note that rotating the phase of $\omega$ is gauge equivalent to an
$SU(2)_F$ flavor transformation). We therefore have a description
of $\mz$ in terms of an ${\bf S}^3$ fibration over the interval,
degenerating to a point at one end and to $\CP^1$ at the other.
This is precisely $\CP^2$.

In summary, as two vortices collide their relative orientations
define a point $p\in\CP^2$. The vortices undergo $90^{\rm o}$ scattering
(and, hence, strings undergo reconnection) unless $p$ coincides
with the special point $\omega=0$ on $\CP^2$. Thus the reconnection
probability $P$ is unity. Note that we did not
assume geodesic motion on the moduli space, and the argument for
reconnection goes through even for strings carrying different
currents in the internal space. Although this may seem surprising,
similar results were observed numerically for Witten's
superconducting strings, resulting in excess charge build up at
the interaction point \cite{supermatz}. 

The behavior of the
vortices passing through the special point $\omega=0$ presumably depends on the
quantity $\zeta=2z/\omega$ as $z\rightarrow 0$. For
$\zeta\rightarrow\infty$, we have seen that the vortices pass
through each other unscathed. We suspect that for other values of
$\zeta$, the vortices undergo scattering an angle less than
$90^{\rm o}$.

\subsection{Reconnection of $U(N)$ Strings}

We now turn to the case of vortices in $U(N)$ gauge theories. The
details are similar to the $U(2)$ case so we shall be brief. The
matrix $Z$ remains $2\times 2$, while $\Psi$ is now a $2\times N$
matrix. Once again we may employ the auxiliary $U(2)$ gauge action
to place $Z$ in upper triangular form:
\be
Z=\left(\begin{array}{cc} z & \omega \\ 0 & -z \end{array}\right)\ \ \ \ \ ,\ \ \ \ \ \
\Psi=\left(\begin{array}{ccc} a_1 &\ldots & a_N \\ b_1& \ldots  & b_N \end{array}\right)
\ee
a choice which is preserved by the remnant $U(1)_1\times
U(1)_2\times \Z_2$ action of equations \eqn{u1} and \eqn{z2}. The
same analysis of the previous section shows that asymptotically,
\be
\tilde{\cal M}_{2,N}\rightarrow \frac{\C\times \CP^{N-1}\times \CP^{N-1}}{\Z_2}
\ee
As we have seen, the question of reconnection boils down to the
$\Z_2$ action which, since it is unchanged, ensures that $z^2$ is
the single-valued coordinate as $z\rightarrow 0$ provided
$\zeta\rightarrow 0$. This time the manifold $\mz$ has complex
dimension $2N-2$, and is defined by the quotient construction,
\be \sum_{i=1}^N|a_i|^2+|\omega|^2=r \ \ \ ,\ \ \ \
\sum_{i=1}^N|b_i|^2-|\omega|^2=r\ \ \ \ ,\ \ \ \
\sum_{i=1}^Na_i\bar{b}_i=0 \label{stiefel}\ee
One can view this space as a fibration over the
interval\footnote{We're grateful to James Sparks for discussions
and explanations regarding these issues.} $|\omega|^2\in[0,r]$. To
ensure that the space $\mz$ is smooth, one must check that spheres
degenerate at either end of the interval, rather than a more
complicated space. For $|\omega|^2\neq 0,r$, we may use the
$U(1)_2$ action to set the phase of $\omega$ to a constant. Then
the constraints \eqn{stiefel} define the Stiefel manifold
$V(2,N)\cong U(N)/U(N-2)$ of orthonormal two-frames in $\C^N$.
Dividing by the remaining $U(1)_1$ action, the fiber over a
generic point is $V(2,N)/U(1)_1$.

At the two end points of the interval some submanifold of this
fiber degenerates. When $|\omega|^2=r$, so $a_i=0$, the
constraints \eqn{stiefel}, together with the $U(1)_2$ action,
ensure that the fiber shrinks to $\CP^{N-1}\cong U(N)/U(N-1)\times
U(1)$. This means that the degenerating cycle at $|\omega|^2=r$ is
\be [U(N-2)\times U(1)]\ /\ [U(N-1)\times U(1)]\ \cong\  {\bf
S}^{2N-3}\ee
Meanwhile, at the other end of the interval, when $\omega=0$, the
full $U(2)$ gauge action is restored. When combined with the
constraints \eqn{stiefel}, this gives rise to the Grassmanian  of
complex two-planes in $\C^N$, which can be described as
$G(2,N)\cong U(N)/U(N-2)\times U(2)$. At this end the degenerating
cycle is
\be [U(N-2)\times U(2)]\ /\ [U(N-2)\times U(1)]\ \cong\  {\bf S}^3
\ee
We therefore find that the cycles that degenerate at either end of
the interval are indeed spheres, and the space $\mz$ is smooth.

In summary, the submanifold $\mz$ is smooth and reconnection
occurs unless the vortices collide over the complex codimension 2
submanifold $\mz\cap \mw\cong G(2,N)$, therefore $P=1$.

\section{Symmetry Breaking and Quantum Effects}

So far we have discussed the situation in which the $SU(N_f)$
flavor symmetry of the model is unbroken. Since global symmetries
are unlikely to be exact in  Nature, in this section we discuss
various mechanisms by which the flavor symmetry is broken and/or
the internal modes on the string are lifted.

\subsection{Symmetry Breaking and Monopole Pair Creation}

We start by introducing explicit symmetry breaking terms into the
Lagrangian. We will present two examples in which the moduli space
of vortices gets lifted, leaving behind $N$ different cosmic
strings, each carrying magnetic flux in a different part of the
gauge group.

The simplest symmetry breaking term is a mass for the scalar
fields $q_i$. After a suitable unitary transformation, we have the
potential,
\be V= \frac{\lambda e^2}{2} 
\Tr \Bigl(\sum_i q_i\otimes q_i^\dagger-v^2\Bigr)^2 +
\sum_im^2_iq_i^\dagger q_i \label{def1}\ee
For $m^2_i\ll \lambda e^2v^2$, the theory still lies in the Higgs
phase, with vacuum expectation value
\be q^a_{\ i}= \sqrt {v^2-\frac{m^2_i}{\lambda e^2}}\ \delta^a_{\
i} \equiv
\mu_i\,\delta^a_{\ i}\ \ \mbox{(no sum over $i$)}\ee \\
The symmetry breaking pattern now becomes,
\be U(N_c)\times SU(N_f)\rightarrow SU(N)_{\rm diag}\rightarrow
U(1)_{\rm diag}^{N-1} \label{sb}\ee
where the first, spontaneous, breaking occurs at the scale
$e^2v^2$, while the second, explicit breaking, occurs at the
scales $m_i$. With the $SU(N)_{\rm diag}$ broken, we can no longer
sweep out a moduli space of vortex solutions as in \eqn{cpn} and
the internal $\CP^{N-1}$ space is lifted. What remains are $N_c$
distinct vortex solutions in which the non-abelian field strength
has non-vanishing component within only one of the diagonal
$U(1)\subset U(N_c)$. Let $B$ denote the adjoint valued magnetic
field in the direction of the strings. Then we may embed an
abelian vortex in the $i^{\rm th}$ $U(1)$ subgroup of $U(N_c)$
with
\be B \sim {\rm diag}(0,\ldots, 1,\ldots,0) \label{flux}\ee
Such a vortex has tension $T_{i} \sim \mu_i^2$. Since each of
these vortices is supported by the same topological quantum
number, $\int \Tr\,B$, only one of these strings is globally
stable; the others may all decay into the string with the lowest
tension. We shall discuss one such mechanism for this decay
shortly.

It is simple enough to alter our model to arrange for all $N_c$
strings to have the same tension. We introduce a new, adjoint
valued, scalar field $\phi$, with canonical kinetic term, and
consider the potential,
\be V=\frac{\lambda e^2}{2}\Tr\Bigl(\sum_i q_i\otimes
q_i^\dagger-v^2\Bigr)^2+\sum_iq_i^\dagger (\phi-m_i)^2 q_i
\label{def2}\ee
This is a variant on the potentials that appear in ${\cal N}=2$
SQCD. Unlike the potential \eqn{def1}, symmetry breaking in the
pattern \eqn{sb} now occurs regardless of the relative values of
$m_i$ and (non-zero) $v^2$.  The unique, gapped, vacuum is given
by
\be q^a_{\ i}=v^2\delta^a_{\ i}\ \ \ \ ,\ \ \ \ \ \phi={\rm
diag}(m_1,\ldots,m_N) \ee
In this theory we again have $N_c$ different vortices, with
magnetic flux \eqn{flux}, but now with equal tension
\eqn{tension}. (Note that if we also include an explicit mass $M$
for the adjoint scalar $\phi$ then symmetry breaking only occurs
for suitably small $M$, but the tensions of the vortices remain
equal).

Both deformations \eqn{def1} and \eqn{def2} result in $N_c$
different types of vortices, each embedded in a different,
orthogonal, $U(1)$ subgroup of $U(N_c)$. This ensures that two
strings colliding with energies $E\ll \Delta m_i$ fall into one of
the two categories discussed in Section 2: either the strings are
of the same type (i.e. inhabit the same $U(1)$) and they
reconnect; or they are of different types, and pass through each
other. Unlike the situation where the strings enjoyed an internal
moduli space, there is no need for fine tuning to make the strings
miss each other: the potential does the job for us. Therefore at
energies smaller than the mass splittings $\Delta m_i$, the
classical probability for reconnection is $1/N$. At energies much
larger than this, the masses are negligible and the probability
increases to unity (at least whenever the moduli space
approximation of the previous section is valid).

\subsection*{\it Confined Monopoles}

In fact, quantum effects give rise to a finite probability for
reconnection even for distinct strings. This can occur if one
string turns into another through the creation of a confined
monopole. Here we give an estimate of the magnitude of this
effect. The presence of confined monopoles, acting like beads on
the cosmic string, may have other interesting cosmological
consequences as explored in \cite{bv,olum,request}.

Strings living in different $U(1)$ subgroups are supported by the
same topological invariant $\int\Tr B$, suggesting that they may
transmute into each other. The change of the string, from one type
to another,  occurs by a kink on the string worldsheet which, from
the four-dimensional perspective, has the interpretation of a
confined magnetic monopole. These monopoles were described in
\cite{tong} and further explored in \cite{amime2,ma}. Similar
objects were previously discovered in $\Z_2$ strings in
\cite{kibhind}. The mass of the kink on the worldsheet is
\be M_{\rm kink} \sim r\Delta m_i \sim
\frac{2\pi \langle\phi\rangle}{e^2}\sim M_{\rm monopole} \ee
which has the same parametric dependence as the mass of the
unconfined magnetic monopole. (In the supersymmetric context the
equality $M_{\rm kink}=M_{\rm monopole}$ is exact).
\begin{figure}[tbp]
\begin{center}
\includegraphics[width=13cm]{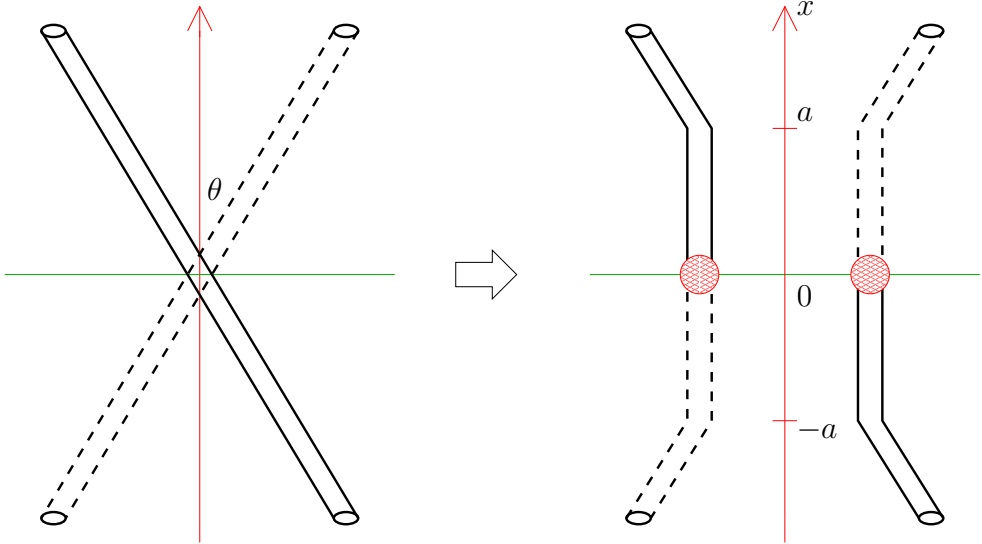}
\put(-70,200){$x$}
\put(-70,160){$a$}
\put(-70,40){$-a$}
\put(-70,90){$0$}
\put(-293,130){$\theta$}
\end{center}
\begin{center}
\begin{minipage}{13cm}
\caption{\small Reconnected strings of different kinds.
The blobs are a monopole and an
 anti-monopole. The dashed lines are original configuration of the
 intersecting strings.}
\end{minipage}
\label{figmonopole}
\end{center}
\end{figure}

Reconnection for different abelian strings requires the quantum
pair creation of a monopole-anti-monopole on the string  as shown
in Figure 3. One can estimate the probability for reconnection to
occur by treating the worldsheet dynamics as a $d=1+1$ quantum
field theory. For simplicity let us model the reconnection of two
almost static strings at incident angle $\theta$ by the shape shown in
Figure 3. Then reconnection reduces the energy of the configuration
by
\begin{eqnarray}
\Delta V = -4 T a \tan^2 (\theta/2) + 2M_{\rm monopole}
\end{eqnarray}
The reconnected region is specified to be $-a < x < a$ where $x$
is the worldsheet spatial coordinate. This is the same potential
arising in electron-positron pair creation in a constant electric
field in $d=1+1$, for which the electric field times the positron charge
is now given by $2T\tan^2(\theta/2)$. The famous result by Schwinger
\cite{schwinger}, evaluating the bounce action of a circular loop
in Euclidean space, gives the decay width as
\begin{eqnarray}
  \Gamma \sim \exp \left(-\frac{\pi M_{\rm monopole}^2}
{2 T \tan^2(\theta/2)} \right)
\sim \exp \left(
-\frac{\pi^2 (\Delta m)^2}{e^4 v^2 \tan^2(\theta/2)}
\right)
\end{eqnarray}
This computation ignores the relative velocity of the strings, and
is valid only for almost parallel strings for which the exponent
is large (and negative). It would be interesting to better
quantify the role of these confined monopoles for other impact
parameters.


\subsection{Quantum Effects}

Until now, much of our discussion has been purely classical.
Indeed, we have chosen the four-dimensional symmetry breaking
scale $e^2v^2$ to be suitably high so the theory is weakly
coupled. Nevertheless, the theory on the vortex string is
necessarily strongly coupled at low-energies: it is the
two-dimensional $\CP^{N-1}$ sigma model.

For now let us set the masses $m_i$ discussed in the previous
section to zero, ensuring that the $SU(N)_{\rm diag}$ symmetry is
exact. The resulting low-energy quantum dynamics on $\CP^{N-1}$ is
well understood. The Mermin-Wagner-Coleman theorem guarantees that
the ground state wavefunction spreads over $\CP^{N-1}$, resulting
in a unique vacuum state for the string. More quantitatively
\cite{amazing}, the size of the vortex moduli space evolves under
RG flow, resulting in dynamical transmutation and a mass gap for
the internal modes on the vortex string. The one-loop beta
function leads to the strong coupling scale
\be \Lambda_{\CP^{N-1}}=\mu \exp\left(-\frac {2\pi
r(\mu)}{N_c}\right)\ee
where, from equation \eqn{r}, we have $r\sim 2\pi/e^2$ at the
symmetry breaking scale $\mu=ev$. Thus the currents discussed
previously, which classically may travel along the worldsheet at
the speed of light, become massive and do not persist. One can
show using the large N expansion that the theory confines and all
dynamical degrees of freedom are singlets of $SU(N)_{\rm diag}$
\cite{amazing}.

In terms of reconnection, the quantum effects do little to change
the story: at low energies $E\ll \Lambda_{\CP^{N-1}}$, the strings
lie in a unique ground state and reconnect with unit probability.
At higher energies, $\Lambda_{\CP^{N-1}}< E < ev$, asymptotic
freedom of the sigma model ensures that the classical analysis of
the previous section is valid and, once again, the strings
reconnect. At energies $E\gg ev$, numerical simulations are
required to determine the issue. Introducing masses $m_i$ as in
\eqn{def1} or \eqn{def2} leads to a weakly coupled theory when
$\Delta m_i\gg \Lambda_{\CP^{N-1}}$ and the results of the
previous subsection hold only in this regime.

\subsection{Fermionic Zero Modes}

The low-energy quantum dynamics of the string can be dramatically
changed by the inclusion of fermionic zero modes \cite{amazing}.
We may add Weyl fermions $\xi$ and $\psi$ to the bulk theory with
Yukawa couplings of the schematic form,
 \be L_{\rm Yukawa} = \bar{\psi}\xi q \label{yuk}\ee
Any such coupling will lead to chiral fermionic zero modes $\chi$
propagating on the vortex string. The exact nature of these zero
mode depends on the properties of $\xi$ and $\psi$ and, as we now
discuss, different color and flavor representations for $\xi$ and
$\chi$ will lead to very different low-energy physics for the
cosmic strings. Here we sketch two examples. More details will be
given in a future publication.

First consider the example in which $\psi_i$ transforms, like
$q_i$, in the fundamental ${\bf N_c}$ of the gauge group, as well
as the fundamental ${\bf N_f}$ of the flavor group (recall that
$N_c=N_f=N$), while $\xi$ is a singlet under both. Then the Yukawa
coupling \eqn{yuk} can be shown to give rise to a single chiral
zero mode on the worldsheet with kinetic term,
\be L_{\rm zero mode} = i\bar{\chi}\delslash\chi \ee
Such zero modes do not couple to $\CP^{N-1}$ modes on the string
and do not affect the low-energy dynamics. (Note that the
four-dimensional anomaly can be cancelled by the addition of
further fermions transforming in the conjugate representation
which may give rise to further fermionic zero modes but do not
qualitatively change the low-energy string dynamics).

A more interesting example comes if we consider $\psi$ to
transform, once again, in the $({\bf N_c},{\bf N_f})$
representation of $U(N_c)\times SU(N_f)$, while $\xi$ transforms
in the adjoint representation of $U(N_c)$ (and is a singlet under
$SU(N_f)$). In this case index theorems ensure the existence of
$N$ zero modes $\chi_i$ on the worldsheet. However, crucially,
they now couple to the strongly interacting $\CP^{N-1}$ sector of
the theory. Let $\pi_i$, $i=1,\ldots, N$ define homogeneous
coordinates on $\CP^{N-1}$, such that $\sum_{i=1}^N|\pi_i|^2=r$,
with $\CP^{N-1}$ obtained after identifying $\pi_i\equiv
e^{i\alpha}\pi_i$. Then the fermionic zero modes on the worldsheet
can be shown to couple to a bosonic $U(1)$ current,
\be L_{\rm current} = i\bar{\chi}_i\,(\pi^\dagger_j
\!\!\stackrel{\ \leftrightarrow}{\delslash}\!\pi_j) \,\chi_i
\label{current}\ee
Once again, the gauge anomaly can be cancelled by the addition of
conjugate fermions in four dimensions. In fact, such action
guarantees that the $d=1+1$ theory on the string is non-chiral,
cancelling a related sigma-model anomaly on the worldsheet \cite{sigmanom}.

So what is the consequence of the interaction \eqn{current}? The
crucial point, as explained in \cite{amazing}, is worldsheet
chiral symmetry breaking. Classically, the $U(1)$ chiral symmetry
acts as $\chi_i \rightarrow e^{i\beta\gamma_5}\chi_i$ while,
quantum mechanically, only a $\Z_{2N}$ is non-anomalous. The
strong coupling dynamics on the worldsheet induces a condensate
for the zero modes, $\langle\chi\chi\rangle \sim
\Lambda_{\CP^{N-1}}$, breaking the discrete chiral symmetry yet
further: $\Z_{2N}\rightarrow \Z_2$. We find a situation similar to
that of Section 4.1, in which the moduli space of vacua is lifted,
now at a scale $\Lambda_{\CP^{N-1}}$, resulting $N$ different
ground states. Recent studies of the low-energy dynamics of these
vortex strings in both supersymmetric theories identify these ground
states with the $N_c$ vortex strings lying in orthogonal, diagonal
$U(1)\subset U(N_c)$ subgroups \cite{amime2}. Once again we have a situation in
which the strings reconnect with probability $P=1/N$ at energy
scales $E\ll \Lambda_{\CP^{N-1}}$, and with unit probability at
higher energies where the sigma-model becomes asymptotically free
and the classical moduli space approximation holds.

\section{Summary and Conclusions}

We have studied the low-energy dynamics of cosmic strings embedded
in a $U(N_c)$ gauge theory with $N_f=N_c\equiv N$ scalars
transforming in the fundamental representation. The cosmic strings
in this theory obtain an internal $\CP^{N-1}$ space in which they
move. We presented a number of deformations which lift this
internal space at a scale $M$, leaving behind $N_c$ types of
vortex string, each embedded in a different diagonal $U(1)\subset
U(N_c)$. Strings of the same type reconnect, while strings of
different types do not interact. In this manner, the classical
probability of two cosmic strings reconnecting at energies $E\ll
M$ is $P=1/N$.

A reconnection probability of $P=1/N$ may also be achieved by
simply considering $N$ decoupled abelian-Higgs models. Our strings
are  distinguished from this trivial case by two effects. Firstly,
even at energies $E\ll M$, the quantum creation of confined
magnetic monopoles may lift the probability above $P=1/N$. We have
only been able to compute this effect in the limit of small
velocity and small angle of incidence where it is negligible. Away
from this regime it may be the dominant contribution to
reconnection and a better understanding of this
process is desired. The second effect occurs at energy scales $E\gg M$ where
the classical probability for reconnection increases to $P=1$, at
least when the moduli space approximation is valid. We find it
interesting that semi-classical magnetic strings in non-abelian
gauge theories remain strongly coupled at large $N_c$, distinguishing
them from their non-perturbative electric counterparts in QCD-like
theories, which are expected to interact with coupling $\sim
1/N_c^2$.

For energies beyond the moduli space approximation we have been
unable to determine the probability of reconnection analytically,
although experience with the abelian-Higgs model suggests that it
may remain unity up to very high energies. It would, of course, be
interesting to develop numerical simulations to extract the
functional dependence of the probability over the full ranges of
impact velocities and incidence angles.

Finally, an important, outstanding problem is to determine how
the scaling of the string network is affected by the presence of 
a threshold scale $M$, and the associated confined monopoles, acting 
like beads on string, which are created after reconnection. One would expect 
monopoles of to also be created by the Kibble mechanism during formation of the 
initial string network. 
It seems plausible that a suitably chosen $M$ may skew the velocity distribution 
of the strings, giving rise to a larger concentration of low-energy strings. 
This would distinguish our non-abelian cosmic strings from others such as 
abelian strings ($P=1$), strongly coupled QCD-like strings 
($P\sim 1/N_c^2$),
weakly coupled fundamental strings ($P\sim g_s^2$)
and D-strings/wrapped D-branes ($P<1$). 
One can only hope that cosmic strings are one day observed, presenting us with
the challenge of deciding between these different possibilities.

\section*{Acknowledgement}
Our thanks to Nima Arkani-Hamed, Nick Dorey, Amihay Hanany,
Sean Hartnoll and James Sparks for many useful discussions. 
K.~H.~is supported in part by the Grant-in-Aid for Scientific Research (\#12440060
and \#15540256) from the Japan Ministry of Education, Science and
Culture, and by funds provided by the U.S.
Department of Energy (D.O.E.) under cooperative research agreement
DF-FC02-94ER40818.  D.~T.~is grateful to the Royal Society for funding.

\end{document}